\documentclass[twocolumn,showpacs,showkeys,preprintnumbers,prb,amsmath,amssymb]{revtex4}
\usepackage[dvips]{graphicx}
\usepackage{dcolumn}% Align table columns on decimal point
\usepackage{bm}% bold math
\usepackage{color}

\newcommand{\refon}{Ref.~\onlinecite}

\newcommand{\aby}{$\sim$}

\newcommand{\be}{\begin{equation} }

\newcommand{\cmsy}{cm$^{2}$}

\newcommand{\ene}{\end{equation}}

\newcommand{\eqr}{Eq.~\ref}

\newcommand{\figr}{Fig.~\ref}

\newcommand{\hcu}{$H_{c2}$ }

\newcommand{\hcuy}{$H_{c2}$}

\newcommand{\jd}{$j_{d}$ }

\newcommand{\rfff}{$\rho_{f}$ }
\newcommand{\rfffy}{$\rho_{f}$}

\newcommand{\rrho}{{$\rho$} }

\newcommand{\tc}{$T_{c}$ }

\newcommand{\tcy}{$T_{c}$}

\begin{document}

%\draft
\preprint{Phys. Rev. B (in press)}
%{Accepted for publication in Phys. Rev. Lett.}
% (M/S \# LX11674)}
%%\title{Free flux flow in the mixed state of a superconductor}
\title{Evaluating free flux flow in low-pinning molybdenum-germanium
superconducting films}
%Unpinned vortex dynamics in molybdenum-germanium films}

\author{Manlai Liang}
\author{Milind N. Kunchur} 
\email[Corresponding author email: ]{kunchur@sc.edu} 
\homepage{http://www.physics.sc.edu/kunchur}
\affiliation{Department of Physics and Astronomy, University of South
Carolina, Columbia, SC 29208}
\author{Jiong Hua}
\author{Zhili Xiao}
\affiliation{Material Science Division, Argonne National Laboratory, 
Argonne, IL 60439}
\affiliation{Department of Physics, Northern Illinois University, 
De Kalb, IL 60115}

\date{Received 27 April 2010; accepted 16 July 2010}
%\today}
%\date{Submitted on 26 October 2008; accepted on 21 January 2009}
%\maketitle
%\vspace{2em}

\begin{abstract}
Vortex dynamics in 
molybdenum-germanium superconducting films were found to 
well approximate the unpinned free limit even at low driving forces. 
This provided an opportunity to empirically establish
the intrinsic character of free flux flow and to test in detail
the validity of theories for this regime 
beyond the Bardeen-Stephen approximation. 
Our observations are in good agreement with the mean-field result 
of time dependent Ginzburg-Landau theory.
%for this primitive regime of flux motion.
%confirm the prediction of theory of 
%of time-dependent Ginzburg Landau (TDGL) theory. 
%Troy and Dorsey (Phys. Rev. B {\bf 47}, 2715, 1993). 
%but do not support the predictions of the microscopic 
%theory of Larkin and Ovchinnikov for this regime.
\end{abstract}

%74.25.F- Transport properties 
%74.25.Op Mixed states, critical fields, and surface sheaths
%74.25.Sv Critical currents
%74.25.Uv Vortex phases (includes vortex lattices, vortex liquids)
%74.25.Wx Vortex pinning (includes mechanisms and flux creep)

\maketitle
%\vspace{1em}
%% Points:-
% negligible pinning; negligible thermal activation (no creep, TAFF,
%etc.); not layered, not 2-D, kigh kappa and hence London (local), etc.
% \rho_n(T) is v. flat
% negligible normal-state MR
% so one of the simplest and most ideal systems
% Why were these non-linear effects not seen in YBCO -- large gap makes 
% relax. time too short to see non-equil. effects?
% Inst. seen is not hot-e (which has clearly identified critical
% parameter field deps. -- verified in YBCO0. But doesnot agree w/ LO
% either -- so LO is basically wrong -- also it give unreasonable T
% dep. for \tau_E
% In YBCO we were in the top 14% of T/Tc

\section{Introduction and background}
The motion of magnetic flux vortices in the mixed state of type II
superconductors is one of the most studied aspects of
superconductors. The most fundamental transport regime of the mixed state
is the state of free flux flow (FFF) where the ``Lorentz'' force on the
vortices is balanced only by the intrinsic viscous drag, without 
any additional interactions arising from pinning, elastic strains,
thermal gradients, etc. 

One of the rudimentary
characteristics expected of the FFF regime (as long as 
$j$ is not large enough to alter the superconducting state) is that the
transport response be Ohmic, so that the flux-flow chordal 
resistivity $\rho_f=E/j$, not
just the differential resistivity $\rho_d=dE/dj$, is constant;  
here $E$ is the electric field and  $j$ is the current density. 
At sufficiently low values of $B$, where each vortex behaves independently,
\rfff is simply proportional to the number density
of vortices and hence to the magnetic field
$B$. In this limit, elementary theories of flux
flow \cite{tinkham,bs,clem} have shown that the proportionality factor is of
the order of $\rho_n/H_{c2}$ resulting in the Bardeen-Stephen
(BS) relation $\rho_f \sim \rho_n B/H_{c2}$, where $\rho_n$ is the
normal-state resistivity and $H_{c2}$ is the
upper critical field  (we set $\mu_0=1$ and use units of tesla for both the 
magnetic field $B$ as well as the magnetizing field $H$). 
As $B$ is increased beyond this low-field limit, additional effects set
in---such as the suppression of the order parameter with increasing 
$B$, changes in the circulating current patterns around the
vortices, et cetera---causing the response to deviate from \rfff $\propto B$.
More advanced theories such as the work by Larkin and Ovchinnikov \cite{lo}
using a microscopic Green-function approach, 
and various works \cite{schmid,caroli,vecris,ullah,dorsey,troy} based on 
time dependent Ginzburg-Landau theory, calculate the detailed
$\rho_f(B,T)$ behavior beyond the Bardeen-Stephen approximation.
These theories have not been sufficiently verified before. 

While there is an impressive body of experimental work that has
discovered or confirmed many interesting and exotic regimes of
vortex dynamics (such as vortex glass, melting, flux creep, vortex
instability, etc.)
% (for an overview and further references please 
%consult the review by Blatter et al. \cite{blatter}) 
the fundamental FFF regime is the least completely studied of these. 
Even the expected Ohmic behavior is usually not observed with high
accuracy; many previous reports  \cite{obs,gapud,dasgupta,lefloch,berghuis} 
either plot the differential ($dE/dj$) instead of the chordal  ($E/j$) 
resistivity, and/or use logarithmic scales so that the deviation from 
Ohmic behavior is less conspicuous than it can be on a linear-linear
plot of \rrho versus $j$. 
 In some earlier work \cite{obs,gapud,dasgupta,lefloch}, 
high current densities were able to largely overcome pinning so 
that $\rho$ became at least roughly constant. 
These previous works were able to establish, at least on a coarse
scale, the adherence to the basic BS behavior. There are a few partial
reports \cite{berghuis,babic-noise,babic-instability} that compare an observed 
$\rho_f(B,T)$ curve with the LO theory, but the FFF regime 
has not been systematically investigated so as to make a detailed
evaluation of the different theories that apply to this regime. 

In the present work we were able to cleanly observe FFF behavior in
low-pinning molybdenum-germanium (MoGe) superconducting films, over
the temperature range $0.6 T_c \alt T \leq T_c$, with exact Ohmic
behavior visible on an uncompressed linear-linear scale even for low
current densities ($j/j_d \sim 1 \times 10^{-4}$ 
and $j/j^* \sim  1 \times 10^{-2}$ where $j_d$ and $j^*$ are the
depairing and flux-flow-instability values respectively) 
that avoid non-linear alterations that 
can arise from high $j$. 
%%% take Hc2(0) ~ 10 T; \lambda=300 nm (Graybeal thesis)
%%% I = 5 uA => j = 5e-6/(6e-6*50e-9) = 16,666,666.67 A/m^2 = 1.7 kA/cm^2
%%% gives: 5.56e-3*sqrt(10)/(3e-7)^2 = 1.953+011 A/m^2 = 2e7 A/cm2
%We especially focus on low to moderate $B/H_{c2}$ range, since the 
%region very close to the \hcut boundary is complicated by 
%fluctuations, cross over to 2 dimensional behavior, 
%homogeneity, etc. 
Our data allow us to empirically elucidate the character of the
FFF regime and to test theories that make detailed predictions
beyond Bardeen-Stephen. 

%On the theoretical front we find that 
All theories for free flux flow are restricted in one way or another (e.g., $T$
close to \tcy , $B$ close to \hcuy, gapless case, etc.). 
%In fact one of the useful outcomes of this work is to experimentally  
%determine an empirical function for $\rho_f(B,T)$. 
A microscopic treatment of this phenomenon,
which is unrestricted in its $B/H_{c2}$ range (although restricted to
$T$ close to $T_c$) is
that due to Larkin and Ovchinnikov \cite{lo} (LO) who find (their Eq.~22):
%That result and its theoretical underpinnings are 
%described in their own book chapter \cite{lo} as well as 
%in the book by Kopnin \cite{kopninbook}. The resulting 
%expression can be written as (their Eq.~22)
\be \sigma_f = \sigma_n + \frac{\sigma_n}{(1-T/T_c)^{1/2}} \frac{H_{c2}}{B}
\tilde{f}(B/H_{c2}), \label{loeq} \ene
where non-linearities that occur at higher values of $B$ are contained
in the $\tilde{f}(B/H_{c2})$
function that they describe in their paper (their Table~1 and
related text). 

Alternatively, various authors 
\cite{schmid,caroli,vecris,ullah,dorsey,troy} have calculated the flux-flow
conductivity using a TDGL (time dependent Ginzburg-Landau) approach.
%Troy and Dorsey \cite{troy} (TD)  use the 
%approach to calculate the
%flux-flow conductivity. %For 
The applicable mean-field result can be written as:
%  (their Eq.~2.37)
\be \sigma_f = \sigma_n + \sigma_n \left( \frac{H_{c2} - 
B}{\nu B}\right), \label{main} \ene
where $\nu \sim 0.4$ is a roughly $T$, $B$, and material independent parameter 
\cite{nucalc}. 
Despite TDGL's strict validity for only gapless
superconductivity, we find that \eqr{main} fits our
observations quite well, taking $\nu$ as an adjustable parameter
that we allow to be determined by the experiment. 

\section{Experimental techniques}
%In the present work, 
Films of thickness $t=50$ nm were sputtered 
onto silicon substrates with 200 nm thick oxide layers from a 
Mo$_{0.79}$Ge$_{0.21}$ alloy target. 
The deposition system had %AJA ATC2400 
a base pressure of $2 \times 10^{-7}$ Torr and 
the argon gas working pressure was maintained at 3
mTorr during the sputtering. The growth rate was 0.15 nm/s.
% as measured by a calibrated quartz-crystal thickness monitor. 
The samples were patterned into bridges of length $l=102 \ \mu$m 
and width $w=6 \ \mu$m using photolithography and argon ion milling. 
Measurements were conducted on four samples A--D. Their 
%%%midpoint
transition temperatures
\tcy, normal-state resistances $R_n$, upper-critical-field slopes 
$H'_{c2}  = d H_{c2}/dT|_{T_c}$, corresponding $H_{c2}(0)$ values
(obtained using the WHH [Werthamer, Helfand, and Hohenberg] 
formalism \cite{whh}), and Ginzburg-Landau
(GL) parameter $\kappa =  3.54 \times 10^4 \sqrt{-\rho_n \mu_o H'_{c2}}$ 
(from Kes and Tsuei \cite{kes1983}) are as follows.
%%%MoGe6:
Sample A: \tcy=5.56 K, $R_n$=555 $\Omega$, 
$H'_{c2}$=-3.13 T/K, $H_{c2}(0)$=12.0 T, and $\kappa$=78. 
%%%MoGe7:  
Sample B: \tcy=5.41 K, $R_n$=555 $\Omega$, $H'_{c2}$=-3.13 T/K, 
$H_{c2}(0)$=11.7 T, and $\kappa$=78. 
%%%MoGe3
Sample C: \tcy=5.01 K, $R_n$=630 $\Omega$, $H'_{c2}$=-3.0 T/K, 
$H_{c2}(0)$=10.3 T, and $\kappa$=77.
%%% nu from R-T=0.23
%%%MoGe4.  
Sample D: \tcy=5.00 K, $R_n$=540 $\Omega$,  $H'_{c2}$=-2.63 T/K, 
$H_{c2}(0)$=9.1 T, and $\kappa$=67. 
%%% nu from R-T=0.305
%%%estimated Hc2 \cite{helfand,whh} using Hc2=0.69*H'c2 gives
%% xxx make table (if space)?

The cryostat was a Cryomech PT405 pulsed-tube
closed-cycle refrigerator that went down to about 3.2 K. It was fitted
inside a 1.3 tesla GMW 3475-50 water-cooled copper electromagnet.
Calibrated cernox and hall sensors monitored $T$ and $B$ respectively. 
The triggers of all measuring instruments 
were synchronized with the pulsed-tube compressor cycle
and instrument measurement windows were set to 1 power line
cycle (16.7 ms \aby 3\% of the compressor cycle) which 
ensured a temperature consistency of \aby 10 mK. 
The main electrical resistivity measurements were 
made using a standard dc four-probe method with an in-house built dc 
current source and voltages measured with a
Keithley model 2000 multimeter. Some extended IV curves were measured
using 0.005 \% duty cycle 20 $\mu$s duration pulses (with 
in-house built pulsed-current source and differential preamplifier, 
and a LeCroy model 9314A digital
storage oscilloscope) to obtain a broader view of the behavior that
includes higher currents up to and beyond the vortex instability; 
however, these high-$j$ measurements are not essential for the analysis and
conclusions of the present work, which pertains to the low-driving force
regime. All data presented here were found to be completely reversible and
showed no hysteresis with respect to cycling of $I$, $B$, and $T$. 
Further details of the measurement techniques have been published
in previous review papers \cite{pbreview,mplb}.

\section{Results and analysis}
\begin{figure}[ht] 
\includegraphics[width=0.8\hsize]{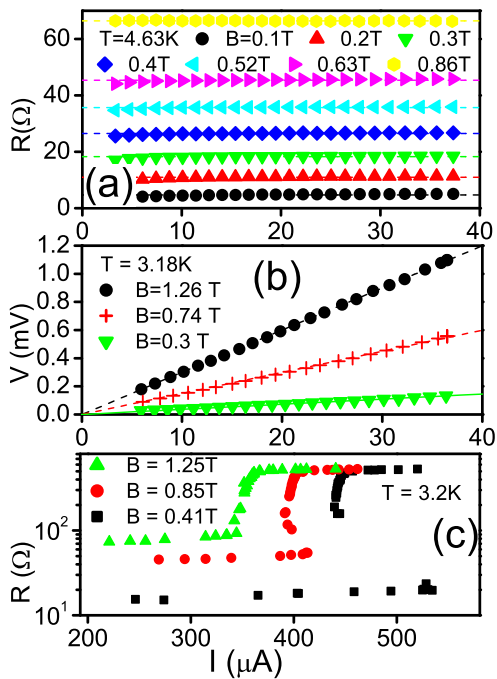}
%%%Z:\Documents\mydocs\PAPERS\MoGe2008\MoGe7\SmallDCT\
%%%SmallDCT-MoGe7-w-IV-plots.OPJ
\vspace{-1em}
\caption{{\em (a) $R$ vs $I$ curves for sample A.  
%(\tcy=5.56 K, $R_n$=555~$\Omega$). 
The horizontal lines are least-squares fits to the data.
(b)  $V$ vs $I$ curves for sample B. 
%(\tcy=5.41 K, $R_n$=555 $\Omega$). 
The lines are least-squares fits to the data extrapolating 
to the origin. (c)  Extended $V$ vs $I$ curves (measured with pulsed
signals) for sample D that include
the vortex instability region.}} %(\tcy=5.00 K, $R_n$=555 $\Omega$).
\label{ohmic}
\end{figure}

\figr{ohmic} shows some examples of mixed-state transport responses. 
Panel (a) shows $R$ vs $I$ curves on a linear-linear
graph to best illustrate the constancy of $R$. 
%(for sample A) 
Panel (b) shows $V$ vs $I$ curves: 
%(for sample B) 
the absence of an intercept again emphasizes that the response is Ohmic
(homogeneously linear) and not just linear with an offset. 
Both panels (a) and (b) were measured with continuous dc currents of
values that are low compared with \jd (by $10^{-4}$) and $j^*$ (by $10^{-2}$).
%but still much higher (by $10^{3}$) than in the work by Babic et al. 
%\cite{babic-noise} who comment on the validity of \eqr{loeq} 
%(they notice effects of pinning and non-Ohmic behavior over 
%a large part of their $B$ range). 
Panel (c) shows an example of a global view of the transport response
%(for sample D) 
that includes the high-$j$ behavior measured using pulsed currents. 
As can be seen, the response becomes progressively more non-linear
with rising $j$, eventually leading to an instability where 
the curves bend around and enter a negative sloped region: There is no 
reentrant Ohmic behavior  (e.g., please see Fig.~2 of 
Ref.~\onlinecite{mplb} or Fig.~5 of Ref.~\onlinecite{blatter}) that
one often sees when the low-$j$ Ohmic region is due to 
thermally assisted flux flow \cite{taff} (TAFF)
 rather than FFF. TAFF, well known from 
high-temperature superconductors, corresponds to the 
case of a weakly pinned flux liquid in the
presence of large thermal activation (when the pinning energy $U_0 \ll k_BT$).
TAFF is characterized by an exponentially reduced resistivity (as in the case
of \refon{babic-noise} where they see an approximately Ohmic response in
the pinned regime at feeble current densities [10 A/\cmsy],  
albeit with a value that is 8\% of \rfffy ) whereas here we 
find \rfff of the same order as the nominal Bardeen-Stephen value
(indeed we find rather detailed agreement with \eqr{main}). 
When a resistive transition is characterized by TAFF, it is expected to show
the Arrhenius dependence $\log R \sim - 1/T$; as we show later our
resistive transitions do not exhibit this behavior but instead follow the
prediction of \eqr{main}. This absence of TAFF behavior is consistent with 
%not surprising 
%for our current densities (which are low compared 
%with  $j^*$ and \jdy, but $10^3$ times
%those of \refon{babic-noise}) 
%since in MoGe, unlike HTS,  
%the flux-lattice melting line $H_m(T)$ is not far below 
%the $H_{c2}(T)$ line because of 
MoGe's small Ginzburg number \cite{ginzburg,blatter} of 
$Gi = 10^{-13} \kappa^4T_c^2/H_{c2}(0) \sim 5 \times 10^{-6}$. 
(In \refon{babic-noise}, in the comparable NbGe system, they observe 
quasi-Ohmic behavior presumably corresponding to TAFF; but their
current densities are $10^3$ times smaller than ours.)
All of the $\rho_f$ data that are subsequently plotted
and analyzed, were checked for Ohmic behavior and
constancy of $R$ as in \figr{ohmic}(a), to ensure that the data lie in
the FFF regime. 

\begin{figure}[ht] 
\includegraphics[width=1.0\hsize]{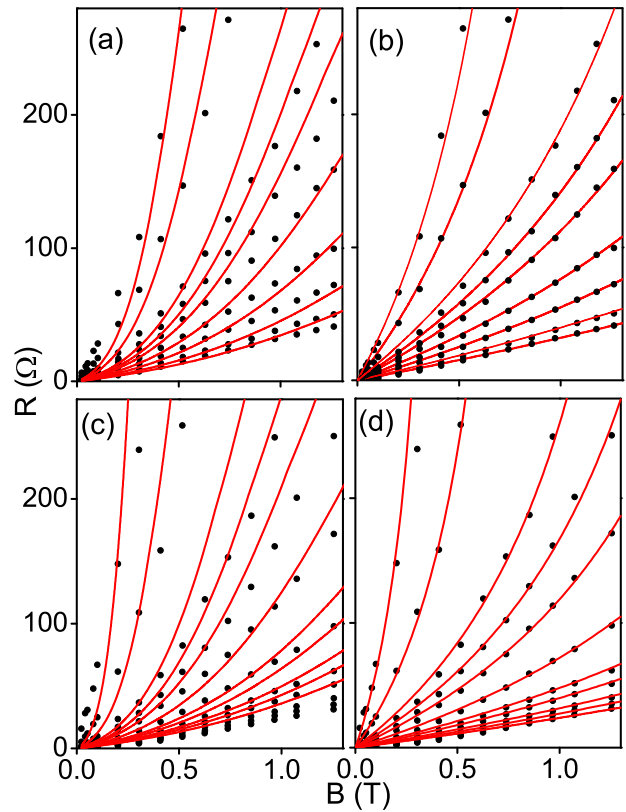}
%%%(a) From "MoGe6RvsBVaryingT-PB-fits-fixedFun-2-with-extended R 
%%%range and both samples.OPJ" from subdir:
\vspace{-1em}
\caption{{\em (a) and (b) 
Resistance versus magnetic field data (symbols) for sample A at
temperatures (bottom to top): 
$T$=3.25, 3.74, 4.24, 4.56, 4.79, 4.90, 5.00, 5.21, 5.31 K. 
%%%T= 3.252, 3.744, 4.235, 4.556, 4.794, 4.901, 5.004, 
%%%5.212,5.308 K. 
(c) and (d) Resistance versus magnetic field data (symbols) for sample B at
temperatures (bottom to top): 
$T$ = 3.17, 3.48, 3.71, 4.01, 4.21, 4.53, 4.77, 4.88, 4.98, 5.19, 5.30 K.
%%%T=3.169,3.483,3.71, 4.013,4.208,4.528, 4.765,4.875,4.981, 
%%%5.188,5.296 K. -- more exactly
(a) and (c) show fits to LO theory (\eqr{loeq}). 
(b) and (d) show fits to the TDGL theory (\eqr{main}) using $\nu$ as a fitting
parameter for each curve (i.e., for each $T$). 
The resulting values of $\nu$ are shown in
\figr{nufactor}. All fits use the same \hcuy$(T)$ given by 
$H_{c2}(T) = H'_{c2}[T-T_{c0}]$ and the measured value
$H'_{c2}$=-3.125 T/K.}}
\label{RvsB}
\end{figure}

\begin{figure}[ht] 
\includegraphics[width=0.8\hsize]{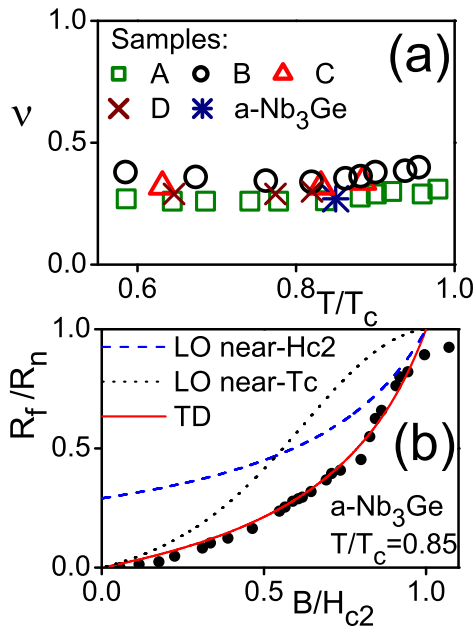}
%%%(a) From "MoGe6RvsBVaryingT-PB-fits-fixedFun-2-with-extended R 
%%%range and both samples.OPJ" from subdir:
%%%Z:\Documents\mydocs\papers\MoGe2008\MoGe6\RvsBManyTs
%%%(b) From a-Nb3Ge-Berghuis-Slot-Kes-PRL 65_2583.OPJ  from subdir:
%%%Z:\Documents\mydocs\PAPERS\MoGe2008\FittingForOtherPapers
\vspace{-1em}
\caption{{\em (a) Experimentally deduced parameter $\nu$ (of \eqr{main})
and its variation with
temperature for MoGe samples A--D. The asterisk shows a
$\nu$ value for the Nb$_3$Ge data of reference
\onlinecite{berghuis} plotted in panel (b). 
(b) Normalized resistance versus normalized field for 
 amorphous Nb$_3$Ge films from Berghuis et al. \cite{berghuis}. 
$t = T/T_c = 0.85$ with \tcy=2.93 K. Solid red line represents a TDGL curve
(\eqr{main}) with
$\nu$=0.27. Blue dashed and black dotted lines correspond 
to the LO theory for the condition 
``close to \hcuy'' (\eqr{lohiB}) with $\tilde{\alpha}$=2.44 and 
for the condition ``close to \tcy''(\eqr{loeq}) respectively.}}
\label{nufactor}
\end{figure}

\figr{RvsB} shows, for samples A (top row) and B (bottom row), the 
free-flux-flow resistances as measured above 
plotted against the magnetic field. The vertical scale corresponds to 
a normalized resistance range of $0 \alt R/R_n \alt 0.5$. (In a later figure
 we show resistive transitions, which
include the $R \rightarrow R_n$ region.)
The different curves correspond to different
temperatures as indicated. At low $B$ the response is linear and at
higher $B$ it curves upward (the simple BS formula corresponds to a linear
response for all $B$). The left column (panels (a) and (c)) shows fits 
to the LO theory (\eqr{loeq}) and the right column  (panels (b) and (d)) 
shows fits to the TDGL theory (\eqr{main}). While both theories are in
principle restricted to work only close to \tcy, the TDGL result provides an
excellent description of the observed behavior over the entire range,  
whereas the LO fits have shapes that do not conform to the
data even for $T/T_c >0.9$ (top 3 curves) where they ought to work 
(it appears that the $\tilde{f}$ function in \eqr{loeq} has 
excessive curvature so that  the LO fits can't
be much improved even with a scaling constant). 
Both theories show a greater departure when  $R \rightarrow R_n$, a region 
that is better revealed in plots of the resistive
transition shown in a later figure. 
For the TDGL fits (panels (b) and (d) of \figr{RvsB}), 
 $\nu$ was taken as a fitting parameter for each curve so 
that the dependence $\nu(T)$ could be determined. This procedure was
repeated for the other two samples. 
 The resulting values of $\nu(T)$ for all four
samples are shown in \figr{nufactor}(a). 
 The  observed $\nu(T)$ is relatively flat 
with respect to temperature with values in the 0.2--0.4 range, which
are consistent with the estimate of $\nu \sim 0.4$ from theory
\cite{nucalc}. 

In the literature, we found one comparable $R$ versus $B$ curve 
%that seems to lie in the FFF regime (even though $R_d$ and 
%not \rfff is plotted) 
for another low pinning system 
(amorphous Nb$_3$Ge) by Berghuis et al. \cite{berghuis} 
(even though they plot differential rather than the chordal resistivity, 
their data appear to lie approximately in the FFF regime).  
Their data are shown in \figr{nufactor}(b).  
As can be seen, the solid red line corresponding to 
the TDGL function (with $\nu=0.27$) fits the entire
range of their data well  (this $\nu$ is shown in panel (a) as an
asterisk). 
In their paper, the authors fit the data to the LO prediction for
the $B \approx H_{c2}$ region (Eq.~30 of LO): 
\be \sigma_f = \sigma_n + \sigma_n \tilde{\alpha} [1- B/H_{c2}]
\label{lohiB} \ene
with $\tilde{\alpha}=2.44$. 
That fit is shown as a dashed blue line in \figr{nufactor}(b) (in their
paper the authors show only the first-order linear expansion $\rho_f/\rho_n
\approx 1 - \tilde{\alpha}[1 - B/H_{c2}]$ of \eqr{lohiB}). 
On the same plot we have also shown a
curve corresponding to \eqr{loeq} (the LO prediction for 
$T \approx T_c$, since these data correspond to $T/T_c = 0.85$) 
as the black dotted line. As can be seen, both LO curves show 
marginal agreement with the data and only over a very narrow
range. 

\begin{figure}[ht] 
\includegraphics[width=1.0\hsize]{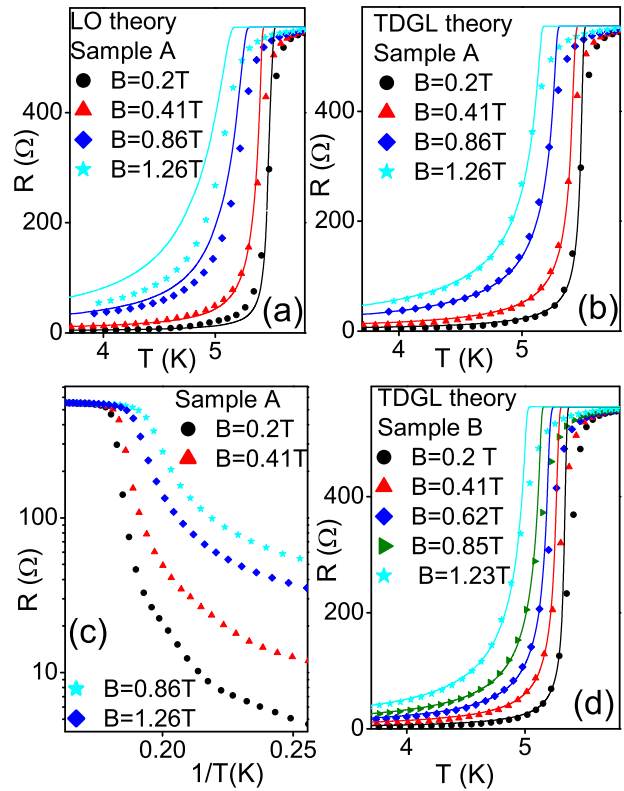}
%%% Origin files are MoGe6R-T-fixed-functions.OPJ from
%%% Z:\Documents\mydocs\PAPERS\MoGe2008\MoGe6\R-T and
%%% MoGe7RTwTc5_41.OPJ from
%%% Z:\Documents\mydocs\PAPERS\MoGe2008\MoGe7\R-T
%%% Final combined Fig is in MoGe6R-T-fixed-functions.OPJ
\vspace{-1em}
\caption{{\em Resistive transitions in magnetic fields. Symbols show
experimental data. Solid lines are theoretical curves. (a) Sample A with
LO theory curves (\eqr{loeq}). (b) Sample A with TDGL theory curves
(\eqr{main} with $\nu$=0.34 for all $B$ and $T$). (c) Arrhenius plots of
%%%$\nu$=0.335 for sample B
the same resistive-transition data for Sample A. (d) 
Sample B with TDGL theory curves
(\eqr{main} with $\nu$=0.26 for all $B$ and $T$).
%%%$\nu$=0.263 for sample B
}}
\label{transitions}
\end{figure}

To explore the region close to $T \approx T_c$ and 
$R_f \approx R_n$, we show in \figr{transitions} the resistive transitions
in various magnetic fields. Panel (a) shows the data for sample A along
with LO curves (\eqr{loeq}); there is some resemblance in trends between
the data and theory, but for the most part the agreement is not close. 
Panels (b) and (d) show the data for samples A and
B along with TDGL curves (\eqr{main}). Given our earlier empirical
confirmation that $\nu$ is $T$ independent (\figr{nufactor}(a)), 
we tried to fit all curves for each sample with a single $\nu$ value. 
As can be seen, for each of panels (b) and (d), 
the TDGL theory provides excellent
agreement with the observed behavior over the lower and middle portion 
of the $R_f/R_n$ range
with a single parameter. The obtained values of $\nu=$ 0.34, 0.26, 0.23, and
0.31 (for samples A--D respectively) are quantitatively consistent 
%%%$\nu=0.34, 0.26, .23, .305
with the theoretically estimated $\nu \sim 0.4$. 
The region right around the
transition does not fit either theory well, perhaps because of a cross over
to 2 dimensionality \cite{babic-noise} or additional effects
\cite{kogan-zhelezina} not included in the present theories, as well as
because of the finite width of the transition. Panel (c) plots the data
for sample A as log $R$ versus $1/T$. As can be seen it does not exhibit
the Arrhenius behavior ($R \propto e^{-1/T}$) that characterizes the TAFF
regime. 

\section{Summary and conclusions}
Free flux flow is the most primitive regime of transport in the mixed
state; however, its experimental investigation is normally hindered by 
gross alterations of the dynamics due
to pinning and other complications (dissociation into pancakes
due to a layered structure, exotic gap symmetries, melting/entanglement
due to high temperatures, etc.). The present experiment has
observed clean FFF behavior in one of the simplest
and nearly model superconductors 
(unpinned, isotropic, low-temperature, weak-coupling BCS, etc.). 
This allowed us to look closely at the 
transport characteristics of this regime beyond the simple
Bardeen-Stephen formula $\rho_f \sim \rho_n B/H_{c2}$ and to assess the
applicability and accuracy of the microscopic and TDGL theories. 
We find that the mean-field result arising out of TDGL
%theory of Troy and Dorsey 
provides a much better agreement with the observed 
$\rho_f(B,T)$ function---and is applicable over a wider range
of conditions---than the theory of Larkin and Ovchinnikov. 

Besides the one experimental curve from \refon{berghuis}
(shown in our \figr{nufactor}(b)) where the authors apply the near-\hcu
 LO $\rho_f(B,T)$ function (\eqr{lohiB}, shown as the blue dashed line in
\figr{nufactor}(b)), in the works by Babic et al. on vortex noise
\cite{babic-noise} and vortex instability \cite{babic-instability}, the
authors show (incidentally to the main investigations of those papers) 
partial agreement with the near-\tc LO $\rho_f(B,T)$ function
(\eqr{loeq}); the insets of their Figs.~1 of both \refon{babic-noise} and
 \refon{babic-instability} each show one experimental curve for one
temperature for one sample.
% with a partial fit to theory. 

The present work only evaluates the accuracy of 
the $\rho_f(B,T)$ function in the FFF regime and 
makes no comments on the accuracy and applicability of the LO
theory for other regimes such as the vortex instability.
An important value of a microscopic theory such as LO 
lies in its ability to go beyond phenomenology and make a connection
between measurements and microscopic parameters (such as the extraction
of the inelastic scattering time as carried out in \refon{berghuis}). 

On the other hand, the mean-field result of the TDGL theory
provides a rather excellent account of the observed $\rho_f(B,T)$
function as
seen from Figs.~\ref{RvsB} and \ref{transitions} along with quantitative
agreement with the predicted parameter $\nu$. This is a useful observation,
because in some calculations of exotic effects---such as vortex
instabilities---the BS formula is often used as a starting point. For such
applications, the TDGL formula (\eqr{main}) with $\nu \sim$ 0.2--0.4
may provide a more accurate alternative while retaining most of the
simplicity of the BS result. 

\section{Acknowledgements}
The authors gratefully acknowledge useful discussions with 
James M. Knight, Alan T. Dorsey, Boris I. Ivlev, Alexander V. Gurevich, 
Vladimir G. Kogan,
Lev Bulaevski, David K. Christen, Ernst Helmut Brandt, and P. H. Kes. 
This work was supported by the U. S. Department of Energy
through grant number DE-FG02-99ER45763. 
The sample fabrication work at Northern Illinois University was
supported by the U.S. Department of Energy through grant number
DE-FG02-06ER46334.

\end{document}